\documentclass[showpacs]{revtex4}

\usepackage{graphicx}
\usepackage{dcolumn}
\usepackage{amsmath}
\makeatletter
\def\btt#1{\texttt{\@backslashchar#1}}
\DeclareRobustCommand\bblash{\btt{\@backslashchar}} \makeatother

\begin{document}

\title[]{Generating dynamical black hole solutions}
\author{A. K.~Dawood}
\author{S. G.~Ghosh}
\thanks{E-mail: sgghosh@iucaa.ernet.in}
\affiliation{%
Birla Institute of Technology and Science (BITS), Pilani - 333 031, INDIA}%

\date{\today}
\begin{abstract}
We prove a theorem that characterizes a large family of non-static
solutions to Einstein equations, representing, in general,
spherically symmetric Type II fluid. It is shown that the best
known dynamical black hole solutions to Einstein equations are
particular cases from this family.  Thus we extend a recent work
of Salgado \cite{ms} to non-static case.  The spherically
symmetric static black hole solutions, for Type I fluid, are also
retrieved.
\end{abstract}


\maketitle

\section{Introduction}
One of the most famous conjectures in general relativity is the so
called {\em cosmic censorship conjecture} (CCC) \cite{rp} which
states that, for physically reasonable initial data, space-time
cannot evolve towards a naked singularity, i.e., the space-time
singularity is always hidden inside black holes, indicating that a
far away observer will not be influenced by it. Despite almost 30
years of effort we are far from a general proof of the CCC. In fact,
it turns out that such a theorem is intractable due to complexity of
the Einstein field equations, hence metrics with special symmetries
are used to construct gravitational collapse solutions. As a result,
there are very few inhomogeneous and nonstatic solutions known, one
of them is the Vaidya metric.

The Vaidya metric \cite{pc}, which has the form
\begin{equation}\label{vm}
ds^2 = - \left[1 - \frac{2 m(v)}{r}\right] d v^2 + 2 \epsilon d v
d r + r^2 (d \theta^2+ sin^2 \theta d \phi^2), \hspace{0.5in}
\epsilon \pm 1
\end{equation}
is a solution of Einstein's equations with spherical symmetry for
a null fluid (radiation) source described by energy momentum
tensor $T_{ab} = \psi l_a l_b$, $ l_a$ being a null vector field.
For the case of an ingoing radial flow, $ \epsilon = 1$ and $m(v)$
is a monotone increasing mass function in the advanced time $v$,
while $\epsilon = -1$ corresponds to an outgoing radial flow, with
$m(v)$ being in this case a monotone decreasing mass function in
the retarded time $v$. The Vaidya's radiating star metric is today
commonly used for two purposes: (i) As a testing ground for
various formulations of the CCC. (ii) As an exterior solution for
models of objects consisting of heat-conducting matter. Recently,
it has also proved to be useful in the study of Hawking radiation,
 the process of black-hole evaporation \cite{rp1}, and in the
stochastic gravity program \cite{hv}. Also, several solutions in
which the source is a mixture of a perfect fluid and null
radiation have been obtained in later years \cite{ka}.  This
includes the Bonnor-Vaidya solution \cite{bv} for the charge case,
the Husain solution \cite{vh} with an equation of state $P = k
\rho $.  Glass and Krisch  \cite{gk} further generalized the
Vaidya solution to include a string fluid, while charged strange
quark fluid (SQM) together with the Vaidya null radiation has been
obtained by Harko and Cheng \cite{hc}. Wang and Wu \cite{ww}
further extrapolated the Vaidya solution to more general case,
which include a large family of known solutions. Their
generalization comes from the fact that the linear superposition
of particular solutions is also a solution of Einstein's
equations. The Vaidya solution in the brane-world was discovered
by Dadhich and Ghosh \cite{dg}

Recently, Salgado \cite{ms} proved a theorem characterizing a
three parameter family of static and spherically symmetric
solutions (black holes) to Einstein equations by imposing certain
conditions on the energy momentum tensor (EMT) (see also
\cite{vk,rg,id}). His work was extended to higher dimensional
spacetime by Gallo \cite{eg}. However this is obviously not the
most physical scenario and one would like to consider dynamical
black hole solutions, i.e., black holes with non-trivial time
dependence.

In this paper, we consider an extension of Salgado's work, so that
a large family of exact spherically symmetric dynamical black hole
solutions are possible, including it's generalization to
asymptotically de Sitter/Anti-de Sitter (dS/AdS). It turns out
that the family of solutions generated represent generalization of
Vaidya solutions that include most of the known solutions.

\section{The dynamical black-hole solutions}

\noindent {\it{\bf Theorem - I}:  Let ($M,~g_{ab}$) be a four
dimensional space-time [sign$(g_{ab})$ = $(-, +, +, +)$] such that
(i) It is non-static and spherically symmetric, (ii) it satisfies
Einstein field equations, (iii) in the Eddington-Bondi coordinates
where \\ $ ds^2 = - A(v,r)^2 f(v,r)\;  dv^2
 +  2 \epsilon A(v,r)\; dv\; dr + r^2 (d \theta^2+ sin^2 \theta d \phi^2)$,
 the energy-momentum tensor $T^{ab}$ satisfies the conditions
$T^v_v = T^r_r$ and $ T^{\theta}_{\theta} = kT^r_r$, $(k =
\mbox{const.}\; \in \; \mathcal{R}$) (iv) it possesses a regular
Killing horizon or a regular origin. Then the metric of the
space-time is given by}
\begin{equation}
ds^2 = - \left[1 - \frac{2 m(v,r)}{r}\right] d v^2 + 2 \epsilon d
v d r + r^2 (d \theta^2+ sin^2 \theta d \phi^2), \hspace{.3in}
(\epsilon = \pm 1)\label{me}
\end{equation}
where
\begin{equation}
m(v,r) = \left\{ \begin{array}{ll}
        M(v)                     &   \hspace{.5in}    \mbox{if $C(v) = 0$}, \\
         & \\
       M(v) - \frac{4 \pi C(v)}{2k+1}r^{2k+1}   &  \hspace{.5in}     \mbox{if $C(v)
\ne 0$ and $k \neq -1/2$}, \\
         & \\
       M(v) - 4 \pi C(v) \ln r & \hspace{.5in}      \mbox{if $C(v) \neq 0$ and $k =
-1/2$}.
                \end{array}
        \right.                         \label{eq:mv}
\end{equation}
\begin{equation}
T^a_b =  \frac{C(v)}{r^{2(1-k)}} \mbox{{{diag}}}[1, 1, k, k],
\label{emt}
\end{equation}
and
\begin{equation}
T^r_v = \left\{ \begin{array}{ll}
       \frac{1}{4 \pi r^2}\frac{\partial M}{\partial v} -
        \frac{1}{2k+1}\frac{\partial C}{\partial v }r^{2k-1}   &  \hspace{.5in}
        \mbox{if $k \neq -1/2$}, \\
        & \\
       \frac{1}{4 \pi r^2}\frac{\partial M}{\partial v} -
       \frac{1}{r^2}\frac{\partial C}{\partial v }\ln r

       & \hspace{.5in}      \mbox{if $k = -1/2$}.
                \end{array}
        \right.                         \label{emt2}
\end{equation}
\textit{Here, M(v) and C(v) are the integration constants whose
values depend on the boundary conditions and the fundamental
constants of the underlying matter.} \\

\noindent {\bf Proof}: Expressed in terms of Eddington coordinate,
 the metric of general spherically symmetric space-time
 \cite{bi} is,
\begin{equation}
ds^2 = - A(v,r)^2 f(v,r)\;  dv^2
 +  2 \epsilon A(v,r)\; dv\; dr + r^2 (d \theta^2+ sin^2 \theta d \phi^2).
\label{eq:me2}
\end{equation}
Here $A(v,r)$ is an arbitrary function. It is useful to introduce
a local mass function $m(v,r)$ defined by $f(v,r) = 1 - {2
m(v,r)}/{r}$. For $m(v,r) = m(v)$ and $A=1$, the metric reduces to
the standard Vaidya metric. It is the field equation $G^0_1 = 0$
that leads to $ A(v,r) = g(v)$. However, by introducing another
null coordinate $\overline{v} = \int g(v) dv$, we can always set
without the loss of generality, $A(v,r) = 1$. Hence, the metric
takes the form,
\begin{equation}
ds^2 = - \left[1 - \frac{2 m(v,r)}{r}\right] d v^2 + 2\epsilon d v
d r + r^2 (d \theta^2+ sin^2 \theta d \phi^2). \label{metric}
\end{equation}
Therefore the entire family of solutions we are searching for is
determined by a single function $m(v,r)$. Henceforth, we adopt
here a method similar to Salgado \cite{ms} which we modify here to
accommodate the non static case. In what follows, we shall
consider $\epsilon =1$. The non-vanishing components of the
Einstein tensor are
\begin{subequations}
\label{fe1}
\begin{eqnarray}
&& G^r_v   = \frac{2}{r^2} \frac{\partial m}{\partial v},
\label{equationa}
\\
&& G^v_v = G^r_r = - \frac{2}{r^2} \frac{\partial m}{\partial r},
\label{equationb} \\
&& G^{\theta}_{\theta} = G^{\varphi}_{\varphi} = - \frac{1}{r}
\frac{\partial^2 m}{\partial r^2}, \label{equationc}
\end{eqnarray}
\end{subequations}
where $\{ x^a\} = \{v,\;r,\; \theta, \; \varphi \}$. The Einstein
field equations are
\begin{equation}
R_{ab} - \frac{1}{2} R g_{ab} = 8 \pi T_{ab} ,\label{efe1}
\end{equation}
and combining Eqs.~(\ref{fe1}) and (\ref{efe1}), we have if $a
\neq b$, $T^a_b=0$ except for a non-zero off-diagonal component
$T^r_v$. It may be recalled that EMT of a Type II fluid has a
double null eigen vector, whereas an EMT of a Type I fluid has
only one time-like eigen vector \cite{he}. In addition, we observe
that the metric (\ref{metric}) requires that $T_v^v = T_r^r$. Thus
the EMT can be written as :

\[
T^a_b = \left(%
\begin{array}{cccc}
T^v_v \; & 0 \;  & 0 \;  & 0 \; \\

T^r_v & T^r_r & 0 & 0 \\
   0 & 0 &T^{\theta}_{\theta}& 0 \\
   0 & 0 & 0 & T^{\varphi}_{\varphi} \\
\end{array}%
\right).
\]
Enforcing the conservation laws $\nabla_a T^a_b=0$, yields the
following non-trivial differential equations:
\begin{equation}
\frac{\partial T^r_r}{\partial r} = - \frac{2}{r}(T^r_r -
T^{\theta}_{\theta}),  \label{divergence}
\end{equation}
\begin{equation}
\frac{\partial T^v_v}{\partial v} = - \frac{\partial
T^r_v}{\partial r} - \frac{2}{r}T^r_v. \label{divergencea}
\end{equation}
Using the hypothesis that ${ T^{\theta}_{\theta} = kT^{r}_{r} }$,
we obtain the following linear differential equation
\begin{equation}
    \frac{\partial T^r_r }{\partial r} = - \frac{2}{r}(1-k) T^r_r,
      \label{diff}
\end{equation}
which can be easily integrated to give
\begin{equation}
T^r_r = \frac{C(v)}{r^{2(1-k)}}, \label{emts}
\end{equation}
where $C(v)$ is an integration constant. Then, using hypothesis
(iii), we conclude that
\begin{equation} T^a_b =
\frac{C(v)}{r^{2(1-k)}} \mbox{{{diag}}}[1, 1, k, k]. \label{emtll}
\end {equation}
 Now using Eqs.~(\ref{fe1}) and
(\ref{efe1}) and (\ref{emts}), we get ${\partial m}/{\partial r} =
-4 \pi {C(v)}/{r^{-2 k}} $, which trivially integrates to
\begin{equation}
m(v,r) = \left\{ \begin{array}{ll}
        M(v)                      &   \hspace{.5in}    \mbox{if $C(v) = 0$}, \\
         & \\
       M(v) - \frac{4 \pi C(v)}{2k+1}r^{2k+1}   &  \hspace{.5in}     \mbox{if $C(v)
\ne 0$ and $k \neq -1/2$}, \\
        & \\
       M(v) - 4 \pi C(v) \ln r & \hspace{.5in}      \mbox{if $C(v) \neq 0$ and $k =
-1/2$}.
                \end{array}
        \right.
\label{mvr}
\end{equation}
Here the function $M(v)$ arises as an integration constant. What
remains to be calculated is the only non-zero off-diagonal
component $T^r_v$ of the EMT. From Eqs.~(\ref{fe1}) and
(\ref{efe1}), one gets
\begin{equation}
T^r_v = \frac{1}{4 \pi r^2} \frac{\partial m}{\partial v},
\end{equation}
which, on using Eq.~(\ref{mvr}), gives
\begin{equation}
T^r_v = \left\{ \begin{array}{ll}
       \frac{1}{4 \pi r^2}\frac{\partial M}{\partial v} -
        \frac{1}{2k+1}\frac{\partial C}{\partial v }r^{2k-1}   &  \hspace{.5in}
        \mbox{if $k \neq -1/2$}, \\
        & \\
       \frac{1}{4 \pi r^2}\frac{\partial M}{\partial v} -
       \frac{1}{r^2}\frac{\partial C}{\partial v }\ln r
       & \hspace{.5in}      \mbox{if $k = -1/2$}.
                \end{array}
        \right.
\end{equation}
It is seen that Eq.~(\ref{divergencea}) is identically satisfied.
Hence the theorem is proved.

The theorem proved above represents a general class of non-static,
spherically symmetric solutions to Einstein's equations describing
radiating black-holes with the EMT, which satisfies the conditions
in accordance with hypothesis (iii). The static solutions derived
in \cite{ms} can be recovered by setting $M(v) = M, \; C(v) = C$,
with M and C as constants, in which case matter is Type I. The
solutions generated here highly rely on the assumption $(iii)$. On
the other hand, although hypothesis (iv) is not used a priori for
proving the result, but it is indeed suggested by regularity of
the solution at the origin, from which, $ T^v_v= T^r_r|_{r=0}$
(see \cite{ms} for further details).
\begin{widetext}
\begin{table}
\caption{Energy-momentum tensor and the corresponding non-static
space-time they generate alongwith the functions $M(v)$ and
$C(v)$, and the $k$-index associated with each space-time.}
\label{table2}
\begin{ruledtabular}
\begin{tabular}{cccc}
Energy-momentum tensor & Space-time &  Functions: $M(v)$and $C(v)$& $k$-index \\
& & & \\
 \colrule
  &   &  &\\

  $T^a_b = 0,\; T^r_v = \frac{1}{4 \pi r^2} \frac{\partial M}{\partial v}$ & Vaidya & $M(v),\; C(v)=0$ & \\
    &   &  \\
  $T^a_b = \frac{\Lambda}{8\pi}, \; T^r_v=0$ & dS/adS & $M(v)=0, \; C(v)=- \frac{\Lambda}{8 \pi}$ & $k=1$ \\
  &   &  & \\
 $T^v_v=T^r_r= -\frac{a}{8 \pi r^2}$  & Global monopole & $M(v)=0,\; C(v)=-\frac{a}{8\pi}$ & $k=0$ \\
 &   &  & \\
 $T^a_b = - \frac{q^2(v)}{8\pi r^4}$ \mbox{{{diag}}}[1, 1, -1, -1]  &   &  & \\
 $T^r_v = \frac{1}{4 \pi r^3} \left[\frac{r \partial M}{\partial v}-q \frac{\partial q}{\partial v} \right]$
  & Bonnor-Vaidya & $M(v)=f(v),\; C(v)=-\frac{q^2(v)}{8\pi}$ & $k=-1$ \\
     &   &  & \\
 $T^a_b = - \frac{g(v)}{4\pi r^{2(m+1)}}$ \mbox{{{diag}}}[1, 1, -m, -m]  & Husian & $ M(v)=f(v),\; C(v)=-\frac{g(v)}{4 \pi}$  & k = -m \\
 $T^r_v = \frac{1}{4 \pi r^2} \left[\frac{\partial f}{\partial v}-\frac{1}{(2m-1)r^{(2m-1)}}
 \frac{\partial g}{\partial v} \right]$   &   &  \\
      &   &  \\
\end{tabular}
\end{ruledtabular}
\end{table}
\end{widetext}
The family of the solutions outlined here contains, for instance,
Bonnor-Vaidya, dS/AdS \cite{ww}, global monopole \cite{gm},
Husain, Harko-Cheng SQM solution \cite{hc,ns}, and Glass-Krisch
string solutions \cite{gk,sg}. Obviously, by proper choice of the
functions $M(v)$ and $C(v)$, and $k-$index, one can generate as
many solutions as required. The above solutions include most of
the known spherically symmetric solutions  of the Einstein field
equations. Some of the examples of EMT satisfying the conditions
of the theorem and which generates the known space-times are
summarized in Table I.

 The solutions discussed in the section are characterized by two arbitrary
 functions $M(v)$ and $C(v)$, and the cosmological constant  $\Lambda$. Thus one would
like to generalize the above theorem to include $\Lambda$. We can
show that the energy momentum tensor components, in general, can
be written as, $T^a_b = T^a_{(f)b} - \frac{\Lambda}{8
\pi}\delta^a_b$ \cite{ms,eg}, where $\Lambda$ is the cosmological
constant and $T^a_{(f)b}$ is energy momentum tensor of the matter
fields that satisfy $T^{\theta}_{(f)\theta} = kT^r_{(f)r}$. Then
the theorem I can be generalized as:

\noindent {\it{\bf Theorem - II}: Let ($M ,~g_{ab}$) be a four
dimensional space-time [sign$(g_{ab})$ = $(-, +, +, +)$] such that
(i) It is non-static and spherically symmetric, (ii) it satisfies
Einstein field equations, (iii) the total energy-momentum tensor
is given by $T^a_b = T^a_{(f)b} - \frac{\Lambda}{8
\pi}\delta^a_b$, where $\Lambda$ is the cosmological constant and
$T^a_{(f)b}$ is energy momentum tensor of the matter fields, (iv)
in the Eddington coordinates where $ds^2 = - A(v,r)^2 f(v,r)\;
dv^2
 +  2 A(v,r)\; dv\; dr + r^2 (d \theta^2+ sin^2 \theta d \phi^2)$, the
EMT $T^a_{(f)b}$ satisfies the conditions $T^t_{(f)t}=T^r_{(f)r}$,
$T^{\theta}_{(f)\theta} = kT^r_{(f)r}$, $(k = \mbox{const.}\; \in
\; \mathcal{R}$), (v) it possesses a regular Killing horizon or a
regular origin. Then the metric of the space-time is given by
metric (\ref{me})}, where
\begin{equation}
m(v,r) = \left\{ \begin{array}{ll}
        M(v) + \frac{\Lambda r^3}{6}                      &   \hspace{.5in}
\mbox{if $C(v) = 0$}, \\
         & \\
       M(v) - \frac{4 \pi C(v)}{2k+1}r^{2k+1} + \frac{\Lambda r^3}{6}   &
\hspace{.5in}     \mbox{if $C(v) \ne 0$ and $k \neq -1/2$}, \\
         & \\
       M(v) - 4 \pi C(v)  \ln r + \frac{\Lambda r^3}{6}& \hspace{.5in}      \mbox{if
$C(v) \neq 0$ and $k = -1/2$}.
                \end{array}
        \right.                         \label{eq:mvl}
\end{equation}
\begin{equation}
T^a_b =  \frac{C(v)}{r^{2(1-k)}} {\mbox{diag}}[1, 1, k, k] -
\frac{\Lambda}{8 \pi} {\mbox{diag}}[1, 1, 1, 1] \label{emtl}
\end{equation}
and
\begin{equation}
T^r_v = \left\{ \begin{array}{ll}
       \frac{1}{4 \pi r^2}\frac{\partial M}{\partial v} -
        \frac{1}{2k+1}\frac{\partial C}{\partial v }r^{2k-1}   &  \hspace{.5in}
        \mbox{if $k \neq -1/2$}, \\
        & \\
       \frac{1}{4 \pi r^2}\frac{\partial M}{\partial v} -
       \frac{1}{r^2}\frac{\partial C}{\partial v } \ln r
       & \hspace{.5in}      \mbox{if $k = -1/2$}.
                \end{array}
        \right.                         \label{emtl1}
\end{equation}
\textit{Here, M(v) and C(v) are the integration constants whose
values depend on the boundary conditions and the fundamental
constants of
the underlying matter.} \\

Again the conservation law $\nabla_a T^a_b=0$ leads to
\begin{equation}
\frac{\partial T^r_{(f)r}}{\partial r} = - \frac{2}{r}(T^r_{(f)r}
- T^{\theta}_{(f)\theta}).  \label{exdivergence}
\end{equation}
Using the assumption made above that ${ T^{\theta}_{(f)\theta} =
kT^{r}_{(f)r} }$,  we obtain
\begin{equation}
    \frac{\partial T^r_{(f)r} }{\partial r} = - \frac{2}{r}(1-k) T^r_{(f)r},
      \label{exdiff}
\end{equation}
which can be easily integrated to give
\begin{equation}
T^r_{(f)r} = \frac{C(v)}{r^{2(1-k)}}, \label{exemts}
\end{equation}
where $C(v)$ is an integration constant. Then, using hypothesis
$(iii)$ and $(iv)$, we conclude that
\begin{equation}
T^a_b = \frac{C(v)}{r^{2(1-k)}} \mbox{{{diag}}}[1, 1, k, k] -
\frac{\Lambda}{8 \pi} \mbox{{{diag}}}[1, 1, 1, 1].
\label{extotemts}
\end {equation}
 Now using Eqs.~(\ref{fe1}), (\ref{efe1}) and (\ref{extotemts}), we get
\begin{equation}
 \frac{\partial m}{\partial r}
= -4 \pi r^2 \left[\frac{C(v)}{r^{2(1-k)}} - \frac{\Lambda}{8
\pi}\right],
\end{equation} which, on integration, gives, (\ref{eq:mvl}). Next, to calcualte the component $T^r_v$ of the EMT.
 From Eqs.~(\ref{fe1}) and (\ref{efe1}), one gets,
\begin{equation}
T^r_v = \frac{1}{4 \pi r^2} \frac{\partial m}{\partial v},
\end{equation}
which, on using Eq.~(\ref{eq:mvl}), gives ~(\ref{emtl1}). Hence the theorem is proved.\\

\section{Energy Conditions}
The family of solutions discussed here, in general, belongs to
Type II fluid defined in \cite{he}. When $m=m(r)$, we have
$\mu$=0, and the matter field degenerates to type I fluid
\cite{ww}. In the rest frame associated with the observer, the
energy-density of the matter will be given by (assuming $\Lambda =
0$),
\begin{equation}
\mu = T^r_v,\hspace{.1in}\rho = - T^t_t = - T^r_r =
-\frac{C(v)}{r^{2(1-k)}}, \label{energy}
\end{equation}
 and the principal pressures are $P_i =
T^i_i$ (no sum convention).  Therefore $P_r = T^r_r = - \rho$ and
$P_{\theta} = P_{\varphi} = k P_r = -k \rho$ (hypothesis $(iii)$). \\

 \noindent \emph{a) The weak energy
conditions} (WEC): The energy momentum tensor obeys inequality
$T_{ab}w^a w^b \geq 0$ for any timelike vector, i.e.,
\begin{equation}
\mu \geq 0,\hspace{0.1 in}\rho \geq 0,\hspace{0.1 in} P_{\theta}
\geq 0, \hspace{0.1 in} P_{\varphi} \geq 0. \label{wec}
\end{equation}
We say that strong energy condition (SEC), holds for Type II fluid
if, Eq.~(\ref{wec}) is true., i.e., both WEC and SEC, for a Type
II fluid, are identical. \\

\noindent {\emph{b) The dominant energy conditions }}: For any
timelike vector $w_a$, $T^{ab}w_a w_b \geq 0$, and $T^{ab}w_a$ is
non-spacelike vector, i.e.,
\begin{equation}
\mu \geq 0,\hspace{0.1 in}\rho \geq P_{\theta}, P_{\varphi} \geq
0. \hspace{0.1 in}
\end{equation}
Clearly, $(a)$ is satisfied if $C(v)\leq 0, k\leq 0$. However,
$\mu
> 0$ gives the restriction on the choice of the functions $M(v)$
and $C(v)$. From Eq.~(\ref{emt2}), $( k \neq -1/2 ),$ we observe
$\mu
> 0$ requires,
\begin{equation}
\frac{\partial M}{\partial v} - \frac{4 \pi}{(2k+1)}\frac{\partial
C}{\partial v}\hspace{.01in}r^{2k+1} > 0.\end{equation} This, in
general, is satisfied, if
\begin{equation}\label{ecM}
  \frac{\partial M}{\partial v}
> 0, \mbox{ and, either }\frac{\partial C}{\partial v} > 0
\mbox{ and } k < -1/2, \mbox{ or }\frac{\partial C}{\partial v} <
0 \mbox{ and } k > -1/2.
\end{equation}
\noindent On the other hand, for $ k = -1/2 $, $\mu \geq 0$ if
${\partial M}/{\partial C} \geq 4 \pi \ln r$. The DEC holds if
$C(v) \leq 0$  and  $-1 \leq k \leq 0$, and the function $M$ is
subject to the condition (\ref{ecM}).  Clearly,  $0 \leq -k \leq
1$.

\section{Singularity and Horizons}
In this section, we shall discuss the physical properties of the
solutions.  The Ricci $R=R_{ab} R^{ab}$, $R_{ab}$ the Ricci tensor
and Kretschmann invariants($\mbox{K} = R_{abcd} R^{abcd}$,
$R_{abcd}$ the Riemann tensor), for the metric (\ref{me}), reduces
to:
\begin{equation}
R = \frac{128 \pi^2 C(v)^2 (1+k^2)}{r^{4(1-k)}} - \frac{32 \pi
\Lambda C(v) (1+k)}{r^{2(1-k)}} + 4 \Lambda^2 .
\end{equation}
\begin{eqnarray}
K & = &\frac{48 M^2}{r^6} + \frac{256 \pi^2 C(v)^2
}{r^6}\left[{\frac{r^{2k+1}}{2k+1}}\right]^2 [(k-1)^2(2k+1)^2 +
2k(4k-1)] \\ \nonumber && +  \frac{16 \pi M(v)C(v) r^{2k+1}}{r^6
(2k+1)}[1 - (4k-3)^2] + \frac{64 \pi\Lambda C(v)}{3
r^6}\left[\frac{r^{k+2}}{2k+1}\right]^2 \\ \nonumber &&
 \times [(1-4k)+(1-4k^2)(k+2) ] \hspace{.5in} \mbox{if}\hspace{.2in}   k \neq
 -1/2,
 \end{eqnarray}
\begin{eqnarray}
K & = & \frac{48 M^2}{r^6} + \frac{64 \pi C(v)}{r^6}\Big[13\pi
C(v) + M(v)\left[5-6\ln r\right]  \\
\nonumber && - 4\pi C(v)[5-3\ln r ]\ln r\Big] +
\frac{8\Lambda^3}{6} - \frac{32\Lambda C(v)\pi r^3}{3}\hspace{.5
in} \mbox{if}\hspace{.2 in}  k = -1/2.
\end{eqnarray}
These invariants are regular everywhere except at the origin $r =
0$, where they diverge.  Hence, the space-time has the scalar
polynomial singularity \cite{he} at $r=0$.  The nature (a naked
singularity or a black hole) of the singularity can be
characterized by the existence of radial null geodesics emerging
from the singularity.  The singularity is at least locally naked
if there exist such geodesics, and if no such geodesics exist, it
is a black hole. The study of causal structure of the space-time
is beyond the scope of this paper and will be discussed elsewhere.

In order to further discuss the physical nature of our solutions,
we introduce their kinematical parameters. Following York
\cite{jy} a null-vector decomposition of the metric (\ref{me}) is
made of the form
\begin{equation}\label{gab}
g_{ab} = - n_a l_b - l_a n_b + \gamma_{ab},
\end{equation}
where,
\begin{subequations}
\label{nv}
\begin{eqnarray}
n_{a} = \delta_a^v, \: l_{a} = \frac{1}{2} \left[ 1  - \frac{2
m(v,r)}{r} \right ] \delta_{a}^v + \delta_a^r, \label{nva}
 \\
\gamma_{ab} = r^2 \delta_a^{\theta} \delta_b^{\theta} + r^2
\sin^2(\theta) \delta_a^{\varphi} \delta_b^{\varphi}, \label{nvb}
\\
l_{a}l^{a} = n_{a}n^{a} = 0 \; ~l_a n^a = -1, \nonumber \\ l^a
\;\gamma_{ab} = 0; \gamma_{ab} \; n^{b} = 0, \label{nvd}
\end{eqnarray}
\end{subequations} with $m(v,r)$ given by Eq.~(\ref{eq:mv}).  The
optical behavior of null geodesics congruences is governed by the
Raychaudhuri equation
\begin{equation}\label{re}
   \frac{d \Theta}{d v} = \mathcal{K} \Theta - R_{ab}l^al^b-\frac{1}{2}
   \Theta^2 - \sigma_{ab} \sigma^{ab} + \omega_{ab}\omega^{ab},
\end{equation}
with expansion $\Theta$, twist $\omega$, shear $\sigma$, and
surface gravity $\mathcal{K}$. The expansion of the null rays
parameterized by $v$ is given by
\begin{equation}\label{theta}
\Theta = \nabla_a l^a - \mathcal{K},
\end{equation}
where the $\nabla$ is the covariant derivative. In the present
case, $\sigma = \omega = 0$ \cite{jy}, and the surface gravity is,
\begin{equation}\label{sg}
\mathcal{K} = - n^a l^b \nabla_b l_a.
\end{equation}
As demonstrated by York \cite{jy}, horizons can be obtained by
noting that (i) apparent horizons are defined as surface such that
$\Theta \simeq 0$ and (ii) event horizons are surfaces such that
$d \Theta /dv \simeq 0$.  Substituting Eqs.~(\ref{eq:mvl}),
(\ref{nv}) and (\ref{sg}) into Eq.~(\ref{theta}), we get, ($k \ne
-1/2$)
\begin{equation}\label{ah1}
\Theta= \frac{1}{r} \left[1 - \frac{2 M(v)}{r} + Q^2(v) r^{2k} -
\chi^2 r^2 \right],
\end{equation}
where $\chi^2 = \Lambda /3$, and $Q^2(v)={8 \pi C(v)}/{2k+1}$.
Since the York conditions require that at apparent horizons $
\Theta$ vanish, it follows form the Eq.~(\ref{ah1}) that apparent
horizons will satisfy
\begin{equation}\label{ah2}
\chi^2r^3 - Q^2 r^{2k+1} - r + 2M=0,
\end{equation}
which in general has two positive solutions. For $\chi^2 =Q^2= 0$,
we have Schwarzschild horizon $r=2M$, and for $M=Q^2=0$ we have de
Sitter horizon $r=1/ \chi$.  As mentioned above, for $k=-1$, one
gets Bonnor-Vaidya solution, in which case the various horizons
are identified and analyzed by Mallett \cite{rm} and hence, to
conserve space, we shall avoid the repetition of same. For general
$k$, as it stands, Eq.~(\ref{ah2}) will not admit simple closed
form solutions.  However, for
\begin{equation}\label{ahq}
Q^2 = Q^2_c = \frac{-1}{(2 k +1)} \left[
\frac{2k}{2M(2k+1)}\right]^{2 k},
\end{equation}
with $\chi^2 = 0$, the two roots of the Eq.~(\ref{ah2}) coincide
and there is only one horizon
\begin{equation}\label{ahs}
  r = \frac{2 M(2k+1)}{2k}
\end{equation}
  For $Q^2
\leq Q^2_c$ there are two horizons, namely a cosmological horizon
and a black hole horizon.  On the other hand if, the inequality is
reversed, $Q^2 > Q^2_c$ no horizon would form.

\section{Concluding remarks}
In conclusion, we have extended to non-static case a recent
theorem \cite{ms} and it's trivial extension (that includes
cosmological term $\Lambda$), which, with certain restrictions on
the EMT, characterizes a large family of dynamical black hole
solutions, representing, in general, spherically symmetric Type II
fluid. The solutions depend on one parameter $k$, and two
arbitrary functions $M(v)$ and $C(v)$ (modulo energy conditions).
It is possible to generate various solutions by proper choice of
these functions and parameter $k$. Many known solutions are
identified as particular case of this family and hence there
exists realistic matter that follows the restrictions of the
theorem.

The family of solutions discussed here, in general, belongs to
Type II fluid. However, if $M = C = $ constant, we have $\mu$=0,
and the matter field degenerates to type I fluid and we can
generate static black hole solutions by proper choice of these
constants.

A rigorous formulation and proof for either version of CCC is not
available.  Hence, examples showing occurrence of naked
singularities remain the only tool to study the various aspects of
CCC.  However, the lack of exact solutions of Einstein field
equation makes it very difficult.  As a consequence,  we are far
from complete understanding of CCC even in the simple case of
spherical symmetry. The solutions presented here can be useful to
get insights into more general gravitational collapse situations
and in general better understanding of CCC that may help to put
CCC in precise mathematical form.

 \acknowledgements Authors would like to thank IUCAA, Pune for
hospitality while this work was done. The invariants in the
section IV has been calculated using GRTensorII \cite{grt}.

\noindent

\end{document}